\documentclass{iaus}
\usepackage{graphicx}

\title[PN internal kinematics] 
{Planetary Nebula Populations and Kinematics}

\author[Michael G. Richer]   
{Michael G. Richer $^1$
}

\affiliation{$^1$Instituto de Astronom\'\i a, UNAM, PO Box 439027, San Diego, CA 92143\\ 
email: {\tt richer@astrosen.unam.mx} \\[\affilskip]
}

\pubyear{2011}
\volume{283}  
\pagerange{1--8}
\jname{Title of your IAU Symposium}
\editors{A.C. Editor, B.D. Editor \& C.E. Editor, eds.}
\begin{document}

\maketitle

\begin{abstract}
The brightest planetary nebulae achieve similar maximum luminosities, have similar ratios of chemcial abundances, and apparently share similar kinematics in all galaxies.  These similarities, however, are not necessarily expected theoretically and appear to hide important evolutionary differences.  As predicted theoretically, metallicity appears to affect nebular kinematics, if subtly, and there is a clear variation with evolutionary stage.  To the extent that it can be investigated, the internal kinematics for galactic and extragalactic planetary nebulae are similar.  The extragalactic planetary nebulae for which kinematic data exist, though, probably pertain to a small range of progenitor masses, so there may still be much left to learn, particularly concerning the kinematics of planetary nebulae that descend from the more massive progenitors.  
\end{abstract}

\firstsection 
\section{Introduction}

The interacting stellar winds paradigm (\cite{kwoketal1978}) has been very successful in explaining many aspects of the planetary nebula stage of the evolution of low- and intermediate-mass stars.  In this scheme, the evolution of the central star drives the evolution of the nebular shell.  Initially, during the AGB stage, a slow, dense wind removes the envelope of the AGB precursor star, producing a shell that drifts away from the precursor star at $5-15\,\mathrm{km}/\mathrm s$ (e.g., \cite{ramstedtetal2006}).  Once all but a tiny fraction of the AGB envelope has been removed (\cite{paczynski1971}; \cite{schonberner1983}), the star begins to evolve away from the AGB to higher temperature and higher surface gravity at a constant bolometric luminosity (e.g., \cite{schonbernerblocker1993}).  The stellar wind becomes more tenuous and its velocity increases.  This velocity evolution naturally leads to an interaction between these \lq\lq two winds".  Initially, the energy of the shock between the winds can be radiated away (e.g., \cite{kahnbreitschwerdt1990}).  Meanwhile, as the central star's temperature increases, an ionization front eventually propagates into the remnant AGB wind, forming what is observed as the nebular shell.  Since ionization greatly increases the thermal pressure within the nebular shell, the ionization front is accompanied by a shock front.  Beyond some threshold velocity, the energy of the shock between the wind from the central star and the AGB envelope is too great to be radiated away and a contact discontinuity forms (\cite{kahnbreitschwerdt1990}).  A forward shock is driven into the undisturbed AGB envelope upstream from the contact discontinuity and a reverse shock is driven into the free-flowing stellar wind downstream.  The reverse shock creates a region of thermalized shocked stellar wind that radiates in X-rays (e.g., \cite{guerreroetal2000}; \cite{kastneretal2000}) and whose over-pressure creates a hot bubble that works to accelerate and expand the AGB envelope that surrounds it (e.g., \cite{villaveretal2002}; \cite{perinottoetal2004}).  The central star's wind continues to energize this hot bubble until the cessation of nuclear reactions in its hydrogen-burning shell finally causes its luminosity to drop.  

Hydrodynamical models allow the simulation of the entire evolution just described, though the results depend upon details assumed for the different constituents.  Observationally, the AGB wind velocity appears to depend upon metallicity (\cite{woodetal1992}; \cite{ladjaletal2010}), with lower velocity occurring at lower metallicity.  The time evolution of the rate of mass-loss on the AGB determines the density profile in the resulting AGB envelope.  Unfortunately, inferring the density profile from observations is difficult (e.g., \cite{lagadecetal2010}.  Likewise, the details of the time evolution of the central star's temperature, luminosity, mass loss rate, and wind velocity are also important, but not ideally-constrained either at a fixed metallicity or as a function of it.  

Recently, \cite{schonberneretal2010} have computed hydrodynamical models of planetary nebulae including the effect of metallicity on the evolution of the central star.  (The initial AGB envelope structure is held constant, though its metallicity is varied.)  These models reproduce the main results of previous work described above (approximately solar metallicity), but, at lower metallicity, produce thicker and faster expanding nebular shells with smaller central cavities, as a result of the lower nebular cooling and the expected lower wind energy at lower metallicity.  \cite{schonberneretal2010} also publish spatially-integrated line profiles in addition to the usual spatially-resolved profiles (at infinitely high resolution).  The former are important, as they are more easily compared to observations, especially of extragalactic planetary nebulae.  For these reasons, these models are probably the most adequate for comparison with the observed kinematics of planetary nebulae from a wide range of stellar populations.  

\section{Selection criteria and observing technique}

Hydrodynamical models can be compared to two types of observations.  First, and most commonly, models can be compared to the internal kinematics observed in individual objects.  In this case, it is possible to attempt to understand the physical processes, such as jets or flows, that give rise to the observed kinematics (e.g., \cite{garciaseguraetal2006}).  Alternatively, models can be compared to the trends observed in a given population of planetary nebulae.  This type of comparison is more apt to check whether the time evolution predicted by models is reflected in the observed internal kinematics, since the assumption is that the objects are all intrinsically similar, but observed at different times during their evolution.  Here, the focus will be on this second type of comparison.

When comparing the trends or characteristics of the internal kinematics of a population of planetary nebulae, it is necessary to select and observe the population of objects in a similar way.  The selection criteria are particularly important for samples of Galactic planetary nebulae, since the different strengths of the many discovery surveys can bias the results.  Obviously, the poorly-known distances to most individual planetary nebulae do not help.  Clearly, defining and applying criteria designed to select large samples of certain objects is the best defense.  Historically, this has not been done, as the objects have usually been selected on the expectation of interesting kinematics.  Typically, Galactic planetary nebulae are observed in a variety of emission lines, with [N~{\sc ii}]$\lambda$6584 being the most common.  

Selection criteria have usually been more consistently applied in the case of extragalactic planetary nebulae, since detection limits frequently permit only the intrinsically brightest objects to be observed.  Often, extragalactic planetary nebulae are selected according to high luminosity in [O~{\sc iii}]$\lambda$5007 only.  Typically, the internal kinematics of extragalactic objects are observed in the [O~{\sc iii}]$\lambda$5007 line, since it is invariably the brightest line available in the entire population.  

Observations of the internal kinematics of Galactic and extragalactic planetary nebulae often differ in their spatial sampling of the object.  This is equivalent to having a different sampling of the velocity phase space for Galactic and extragalactic planetary nebulae.  Usually, the spectrograph slit intercepts only a fraction of Galactic planetary nebulae and so the observations are spatially-resolved.  As a result, the internal kinematics reflect the motions of only the gas projected within the line of sight defined by the spectrograph slit in the light of the ion observed.  This is extremely useful for studying the fine details of internal motions, especially of faint components such as jets or other structural components that represent a small fraction of the total ionized mass.  For ground-based observations of extragalactic planetary nebulae (to date, these represent \emph{all} observations of their internal kinematics), the objects are point sources at distances beyond the Magellanic Clouds, so the objects in their entirety fit within the slit and the observations lack spatial resolution.  As a result, the internal kinematics of extragalactic planetary nebulae pertain to the entire volume of the emitting ion (usually $\mathrm O^{2+}$).  

\begin{figure}[]
\begin{center}
 \includegraphics[width=0.4\columnwidth]{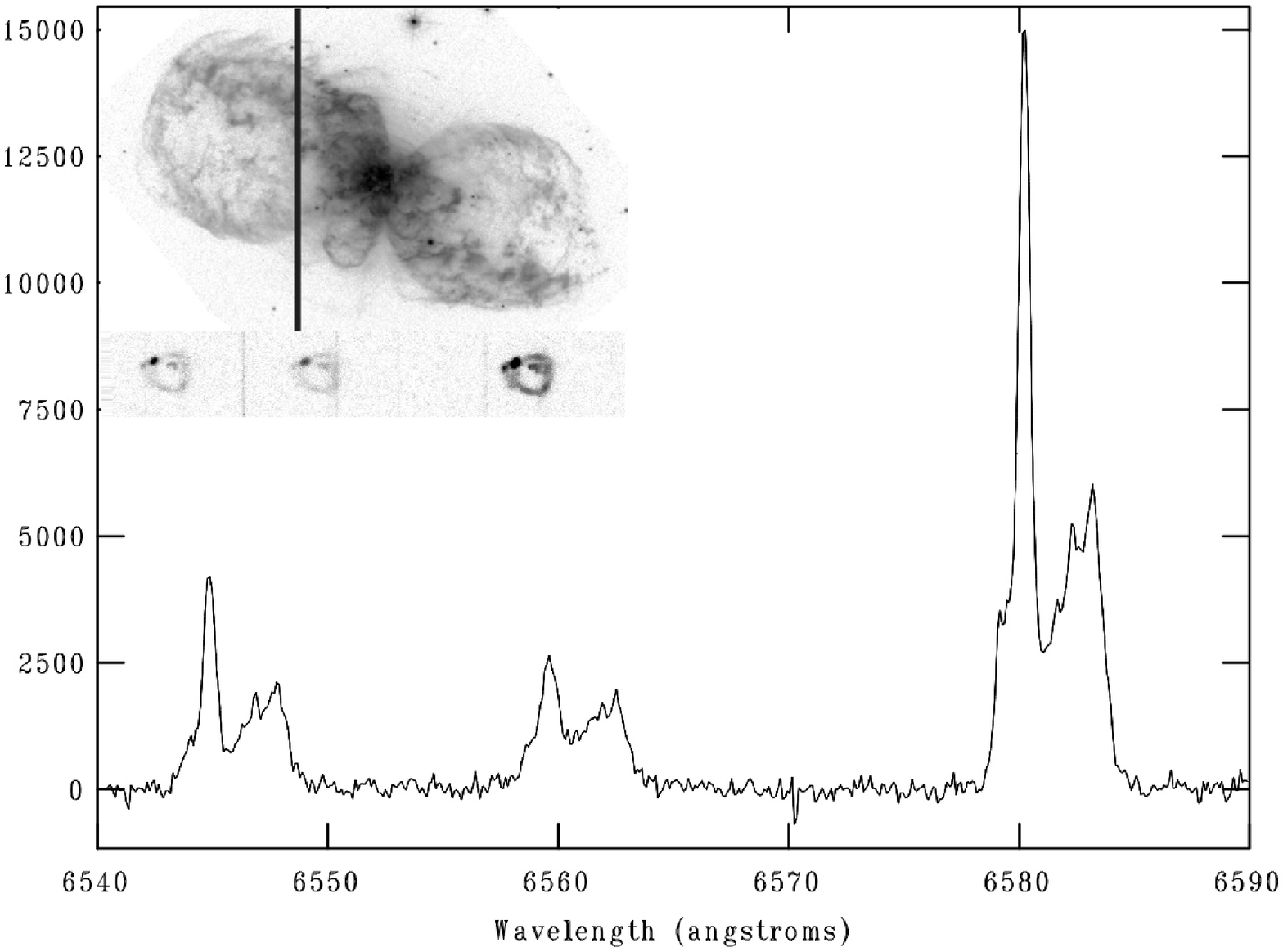} 
 \includegraphics[width=0.4\columnwidth]{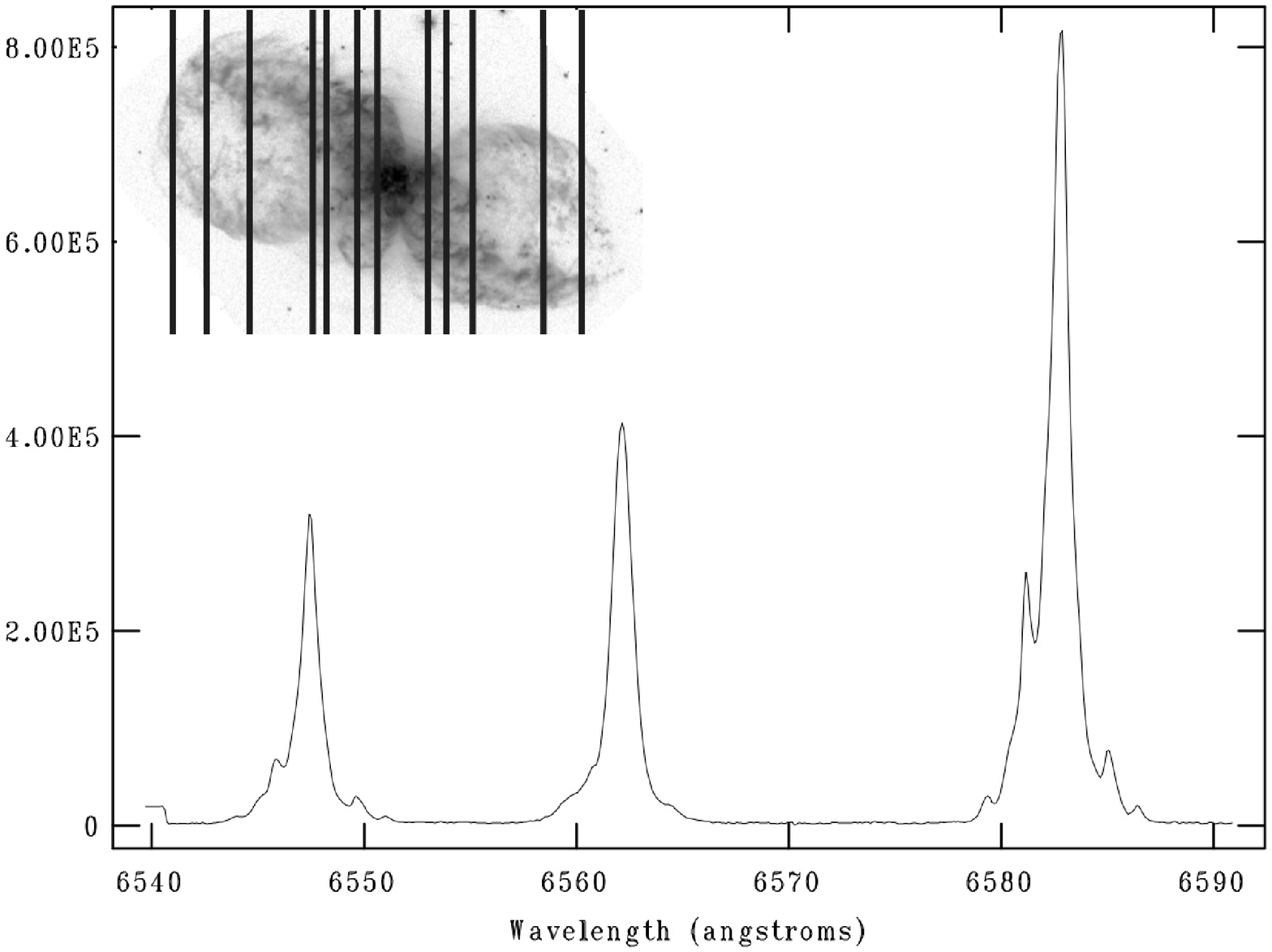} 
 \includegraphics[width=0.13\columnwidth]{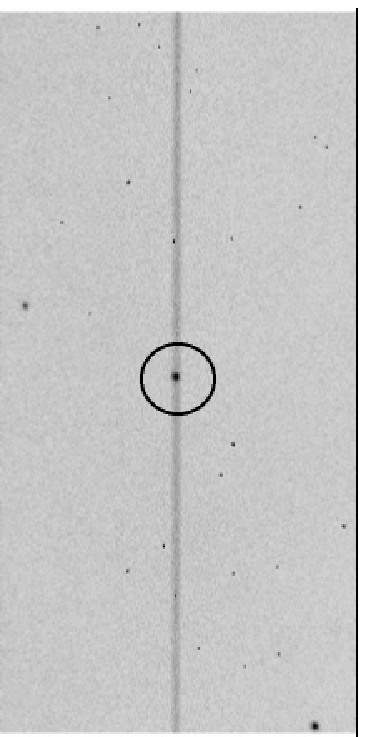} 
 \includegraphics[width=0.4\columnwidth]{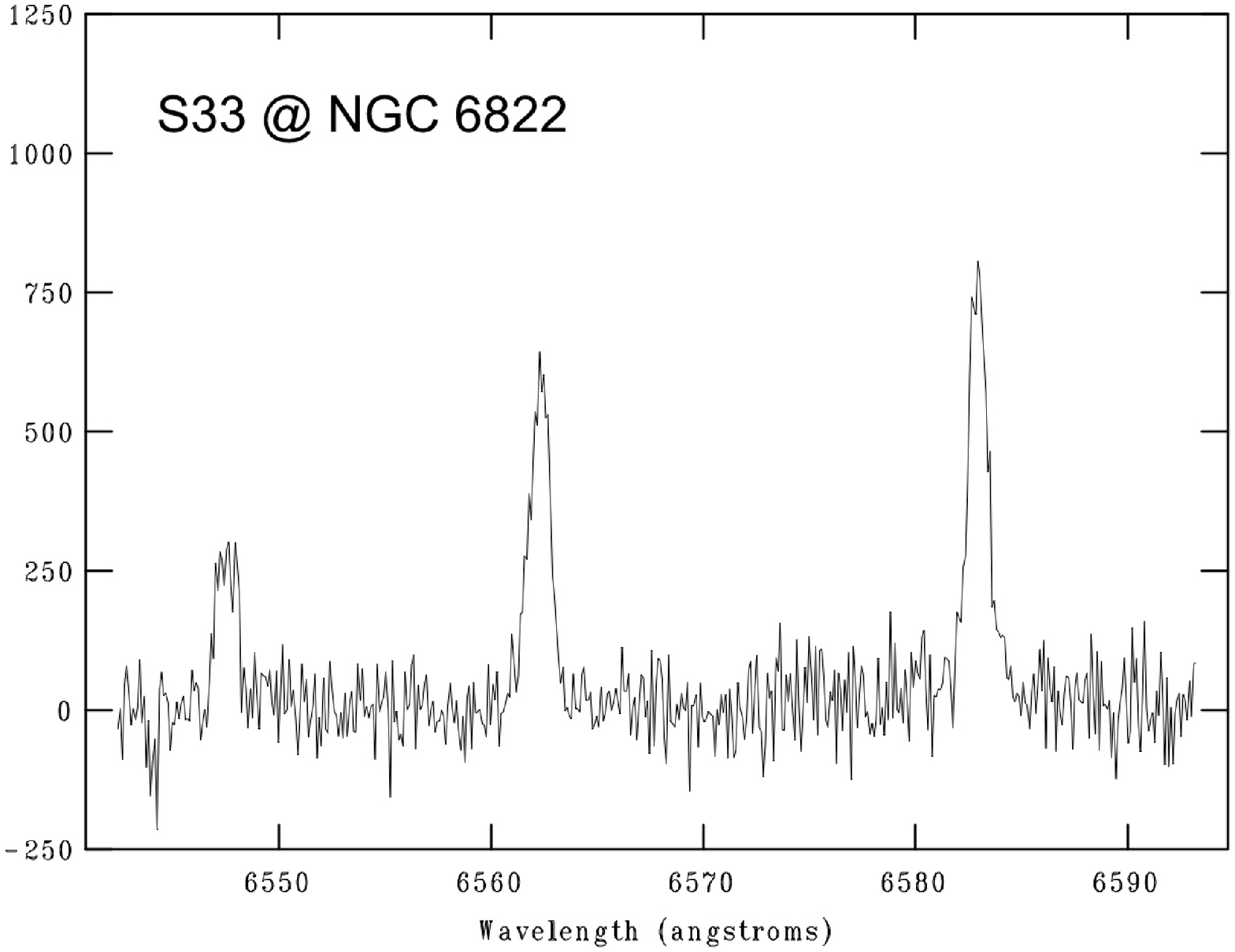} 
 \includegraphics[width=0.4\columnwidth]{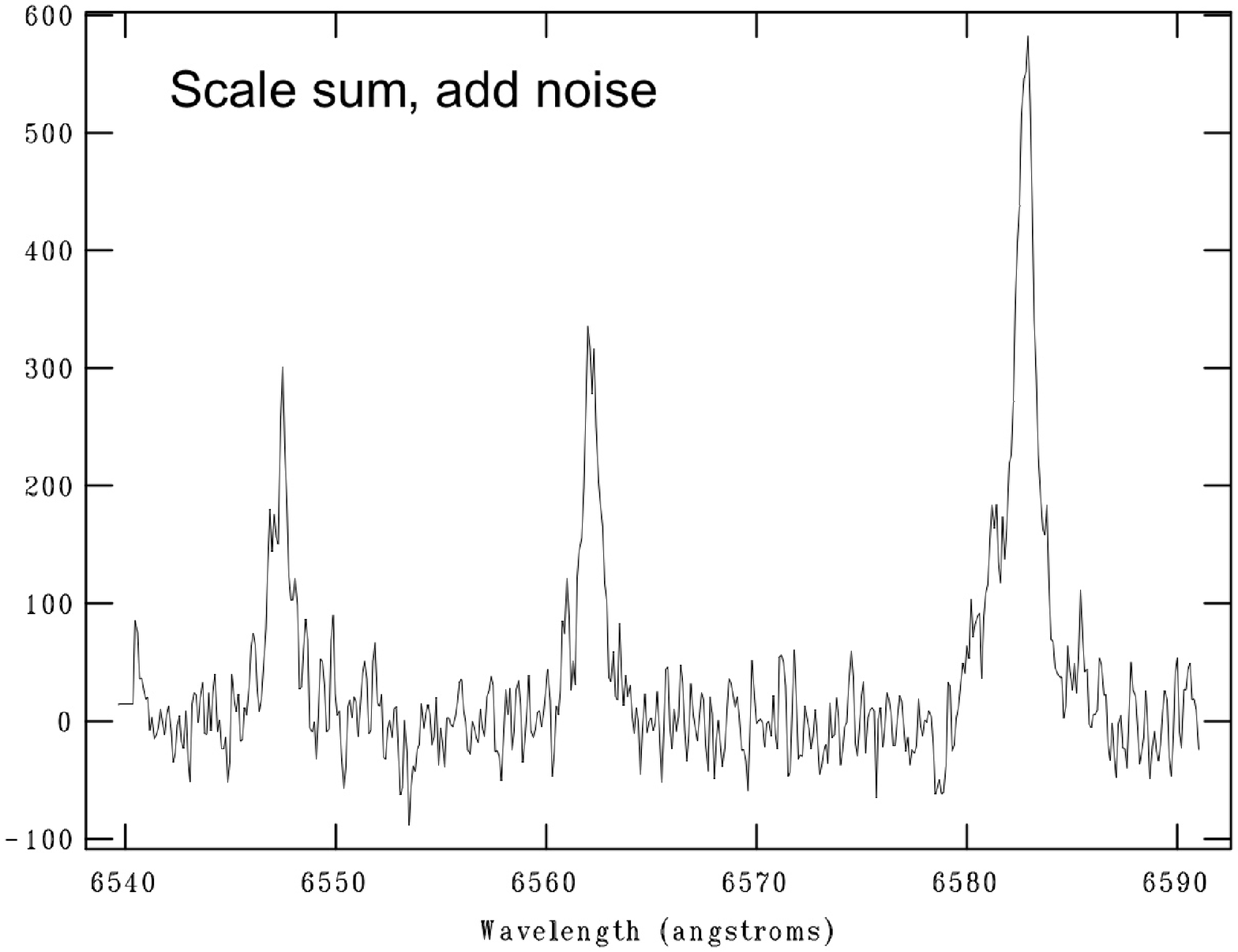} 
 \includegraphics[width=0.18\columnwidth]{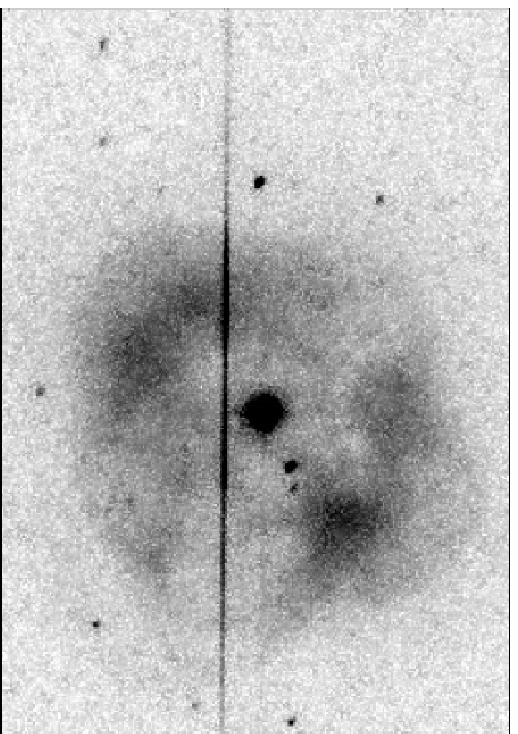} 
 \caption{This figure demonstrates the complementarity of spatially-resolved and spatially-integrated observations.  In the right-most column, examples of the slit superposed upon the planetary nebula in the Fornax galaxy and upon NGC 1514 are shown, illustrating the difference in spatial coverage of the slit.  At top left, the image inset shows a single slit position superposed upon the HST image of Hb 5.  Immediately below, the two-dimensional spectrum of H$\alpha$ and [N~{\sc ii}]$\lambda\lambda$6548,6584 is shown as a second inset.  The main panel at top left shows the result of collapsing this two-dimensional spectrum to a one dimensional spectrum.  This is equivalent to a spatially-unresolved observation of the entire slit.  The panel at top centre shows the effect of collapsing the 12 slit positions shown in the inset image and then summing them.  This is approximately what would be observed in a spatially-unresolved observation of the entire object.  At bottom centre, this sum of 12 spectra has been scaled to the flux typically observed in extragalactic planetary nebulae and added to a typical background.  This can be compared to an observation of the planetary nebula S33 in NGC 6822, shown at bottom left.  The image of Hb 5 is courtesy of Bruce Balick (University of Washington), Vincent Icke (Leiden University), Garrelt Mellema (Stockholm University), and NASA.}
   \label{fig_1d_2d_spec}
\end{center}
\end{figure}

Fig. \ref{fig_1d_2d_spec} demonstrates the complementarity of spatially-resolved and spatially-integrated observations.  In particular, it is evident that deciphering spatially-integrated observations to understand the internal motions in detail is very difficult.  Even estimating the expansion velocity, from the wings of the line profile, is not trivial (\cite{schonberneretal2010}).  The line profiles are well represented by Gaussian profiles, a result in accord with the models of \cite[Sch\"onberner et al. (2010)]{schonberneretal2010}.

\begin{figure}[t]
\begin{center}
 \includegraphics[width=0.49\columnwidth]{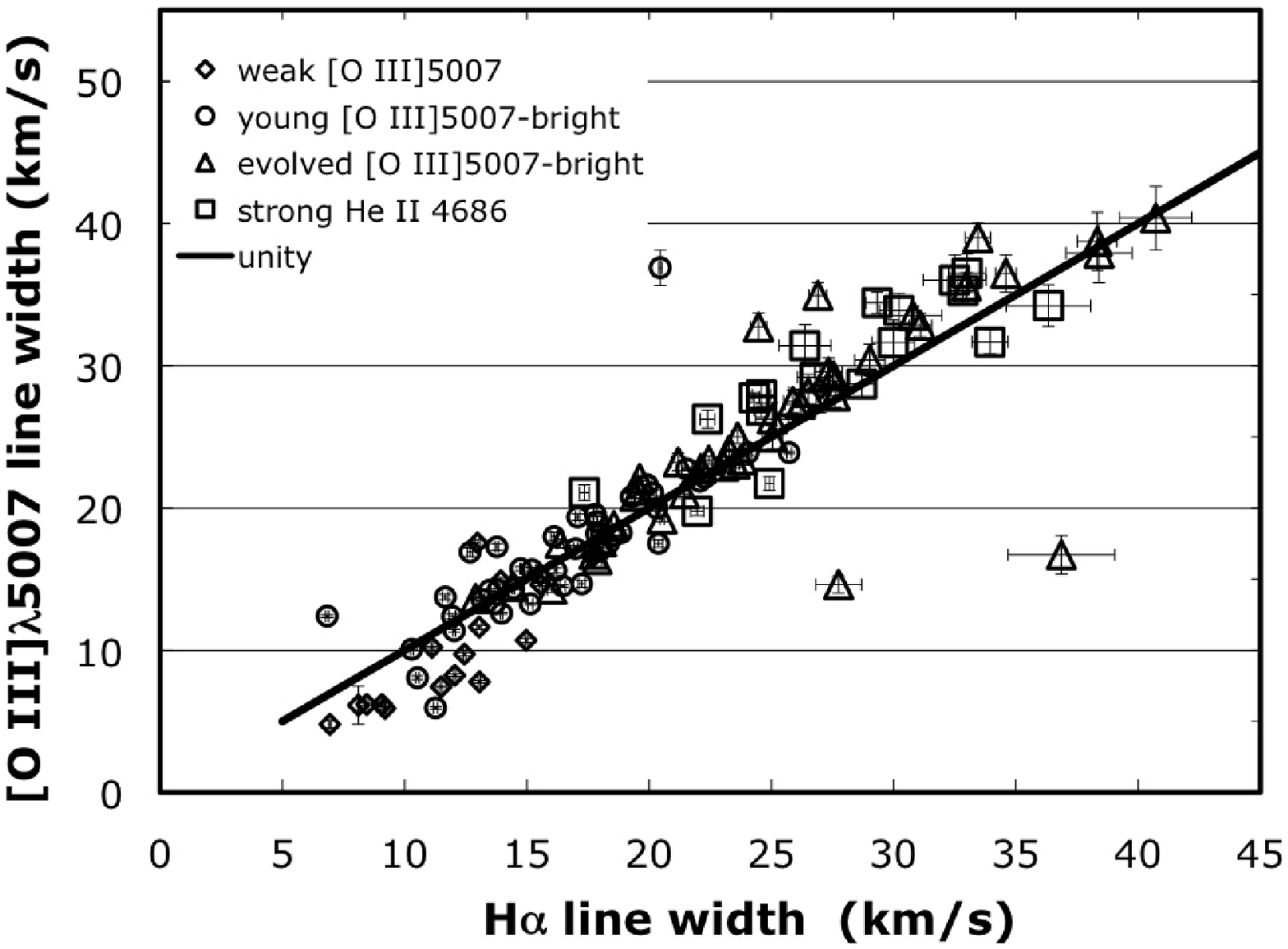} 
 \includegraphics[width=0.49\columnwidth]{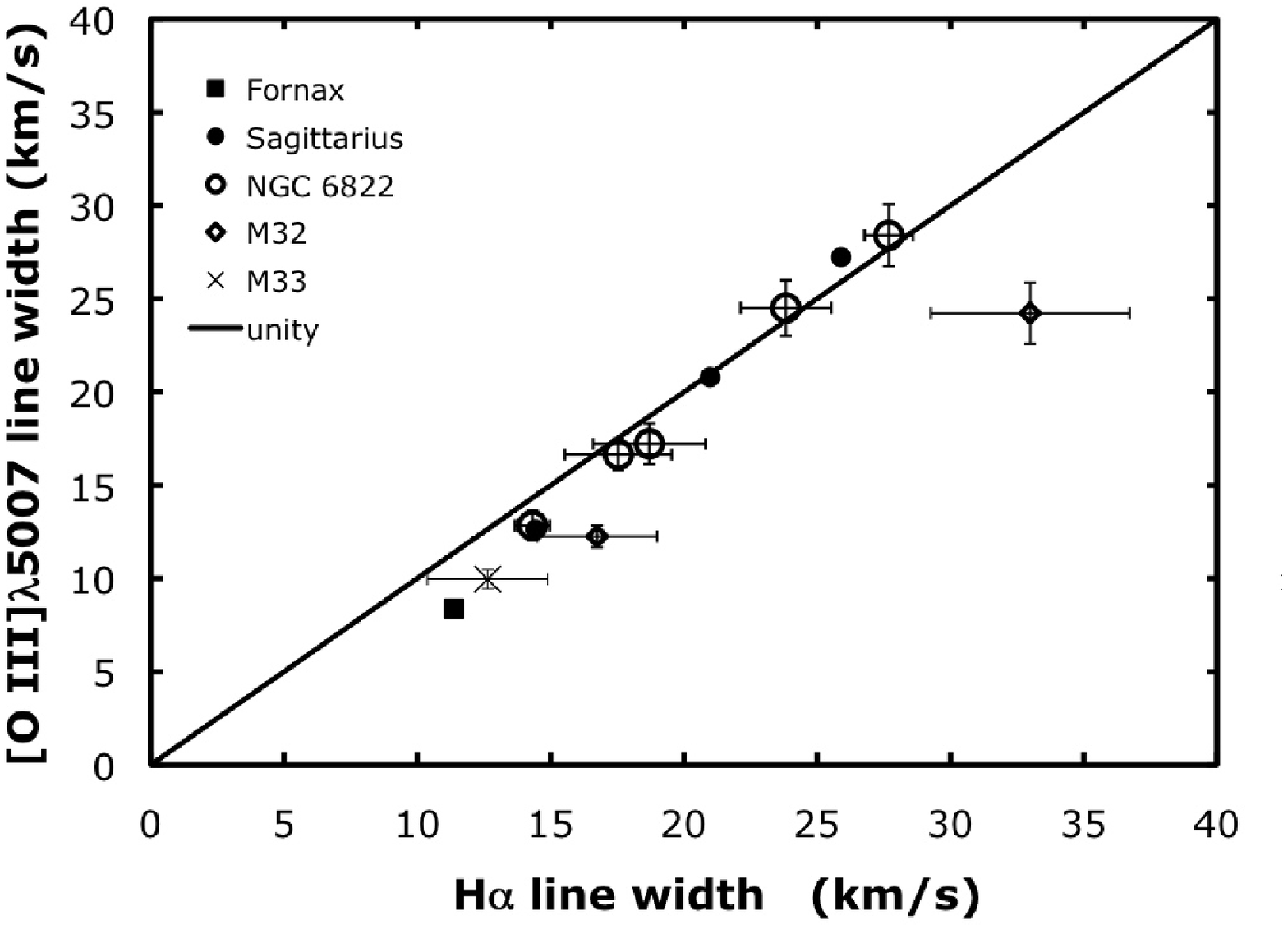} 
 \caption{This figure compares the line widths in H$\alpha$ and [O~{\sc iii}]$\lambda$5007 measured from spatially-integrated observations of planetary nebulae in the Milky Way bulge and in other galaxies (\cite{richeretal2008, richeretal2010a, richeretal2010b}).  The diagonal lines indicate the locus of equal line widths.  None of the planetary nebulae in the galactic bulge are smaller than the spectrograph slit, but the results have been analyzed as if they were (\cite{richeretal2010a}).  Since H$\alpha$ is emitted throughout the entire nebular shell, it is clear that the kinematics in [O~{\sc iii}]$\lambda$5007 is also a reliable indicator of the kinematics of the entire ionized shell (see \cite{richeretal2009} for details).  \cite[Zijlstra et al. (2006)]{zijlstraetal2006} also present kinematic data for the planetary nebulae in the Sagittarius galaxy.}
   \label{fig_ha_o3}
\end{center}
\end{figure}

Finally, Fig. \ref{fig_ha_o3} demonstrates that the kinematics measured from spatially-integrated observations in the lines of H$\alpha$ and [O~{\sc iii}]$\lambda$5007 are equally good indicators of the kinematics of the entire nebular shell for planetary nebulae with intrinsically large [O~{\sc iii}]$\lambda$5007 luminosities.  Given this selection criterion, it is likely that this result arises due to the large volume occupied by the $\mathrm O^{2+}$ ion.  This result is also predicted by the hydrodynamical models of \cite[Sch\"onberner et al. (2010)]{schonberneretal2010}.

\section{Trends in internal kinematics}

\begin{figure}[t]
\begin{center}
 \includegraphics[width=0.6\columnwidth]{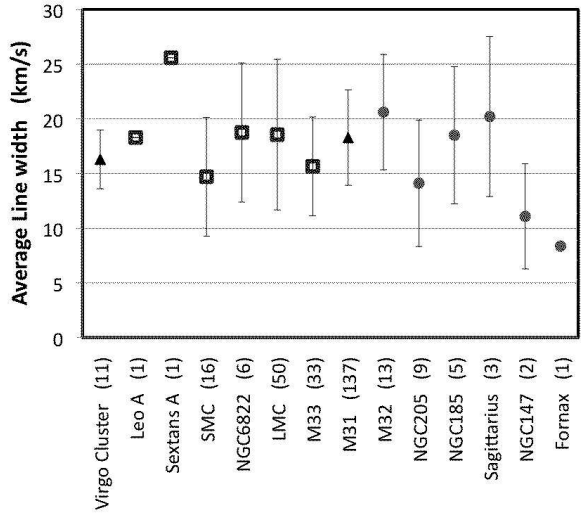} 
 \caption{The average line width measured for intrinsically bright planetary nebulae in a given galaxy (or system) are plotted for 13 galaxies of the Local Group and for intracluster planetary nebulae in the Virgo cluster (\cite{arnaboldietal2008}).  The error bars indicate the standard deviation of the line width distribution in each galaxy.  The line widths are for the [O~{\sc iii}]$\lambda$5007 line and are corrected for instrumental and thermal ($10^4$\,K) broadening.  Galaxies with on-going star formation are to the left (open squares) and those without are to the right (filled circles).  M31 is a mixed system (middle) and it is unknown whether the intracluster planetary nebulae in the Virgo cluster arise from galaxies with or without star formation.  Metallicity is highest at M31 and decreases towards the left and right, i.e., the galaxies with the lowest metallicities are at the edges of the plot.  The data for the planetary nebulae in the Magellanic Clouds are from \cite[Dopita et al. (1985)]{dopitaetal1985} and \cite[Dopita et al. (1988)]{dopitaetal1988}.  Here and throughout, only the intrinsically brightest extragalactic planetary nebulae are considered, within about 2\,mag of the peak of the luminosity function in [O~{\sc iii}]$\lambda$5007.}
   \label{fig_linewidth_galaxies}
\end{center}
\end{figure}

Fig. \ref{fig_linewidth_galaxies} presents the average line width observed for planetary nebulae in galaxies of the Local Group and the intracluster environment of the Virgo Cluster.  The average line width and the range observed does not vary strongly, as a function of environment, metallicity, or the presence of ongoing star formation.  Since the AGB envelope velocities depend upon metallicity (e.g., \cite{woodetal1992}), presumably some mechanism compensates, probably thermal pressure (\cite{schonberneretal2010}).  The lack of a dependence of the line width on the presence of ongoing star formation could indicate that the progenitors are of similar masses in all galaxies, as already indicated by photometry (\cite{ciardulloetal1989}) and chemical abundances (\cite{richermccall2008}).

Hydrodynamical models have long predicted that the internal kinematics evolve with time: the expansion velocity should initially be that of the AGB envelope, it should accelerate as the ionization front is driven through the envelope, and again as a result of the pressure from the hot bubble.  This evolution was first observed for the planetary nebulae in the Magellanic Clouds (\cite{dopitaetal1985, dopitaetal1988}), but its demonstration for planetary nebulae in the Milky Way was plagued by problems related to the distance scale and the small, heterogeneous samples that were used.  Furthermore, a distance- and composition-independent \lq\lq clock" is needed.  In the case of the Magellanic Clouds, the nebular excitation class was used, but the formulation used depends upon the oxygen abundance and so is not universally applicable over a wide range of metallicity (\cite{reidparker2010}).  While no age indicator is likely to be completely independent of distance and composition, the evolutionary state of the central star is perhaps the most natural clock in the context of the interacting stellar winds paradigm, since the central star is expected to drive the evolution of the nebular shell.  During the constant luminosity phase of their evolution, hotter central stars are more evolved.  Once they cease nuclear burning, fainter central stars are more evolved.  

\begin{figure}[t]
\begin{center}
 \includegraphics[width=0.49\columnwidth]{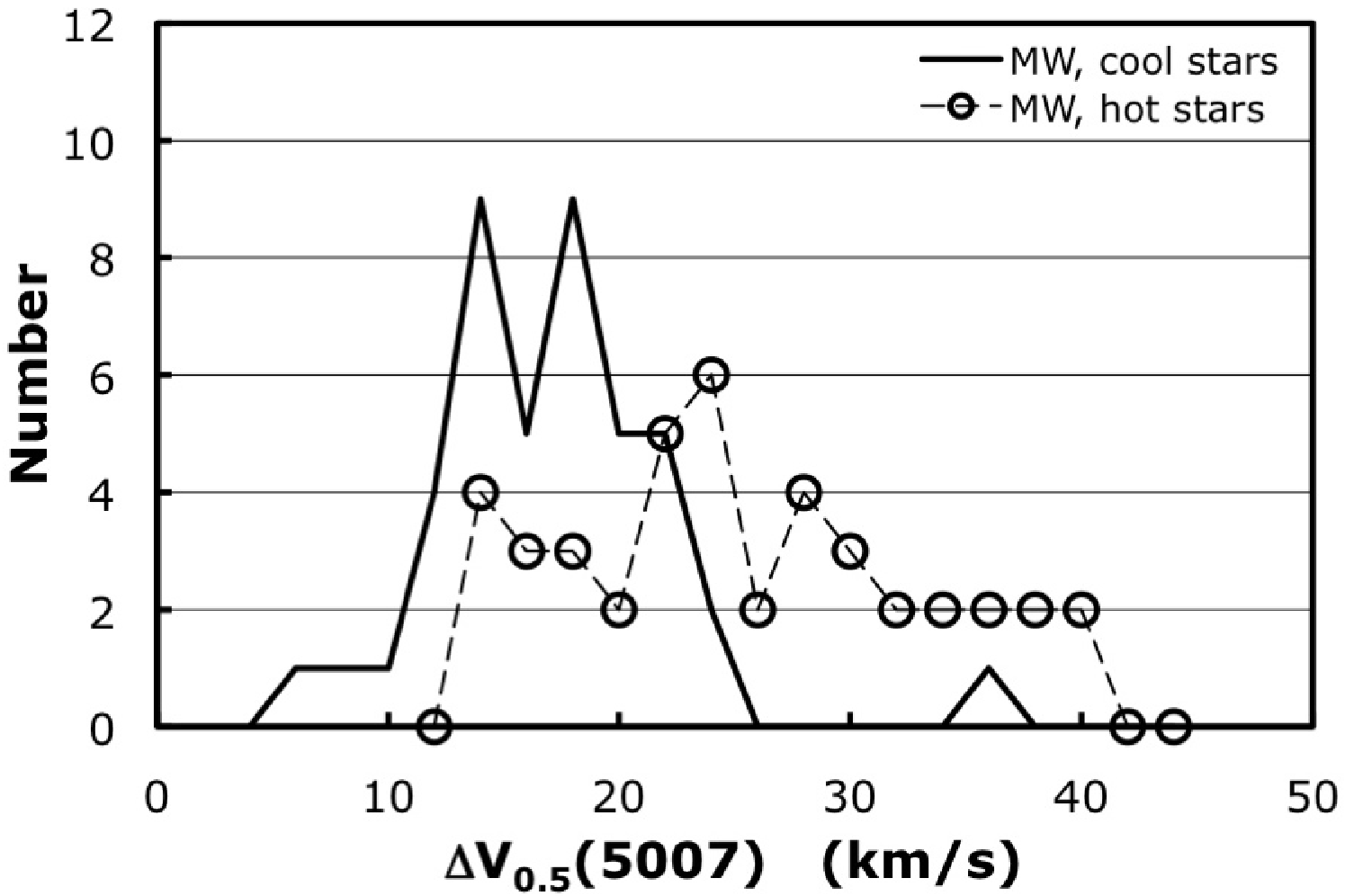} 
 \includegraphics[width=0.49\columnwidth]{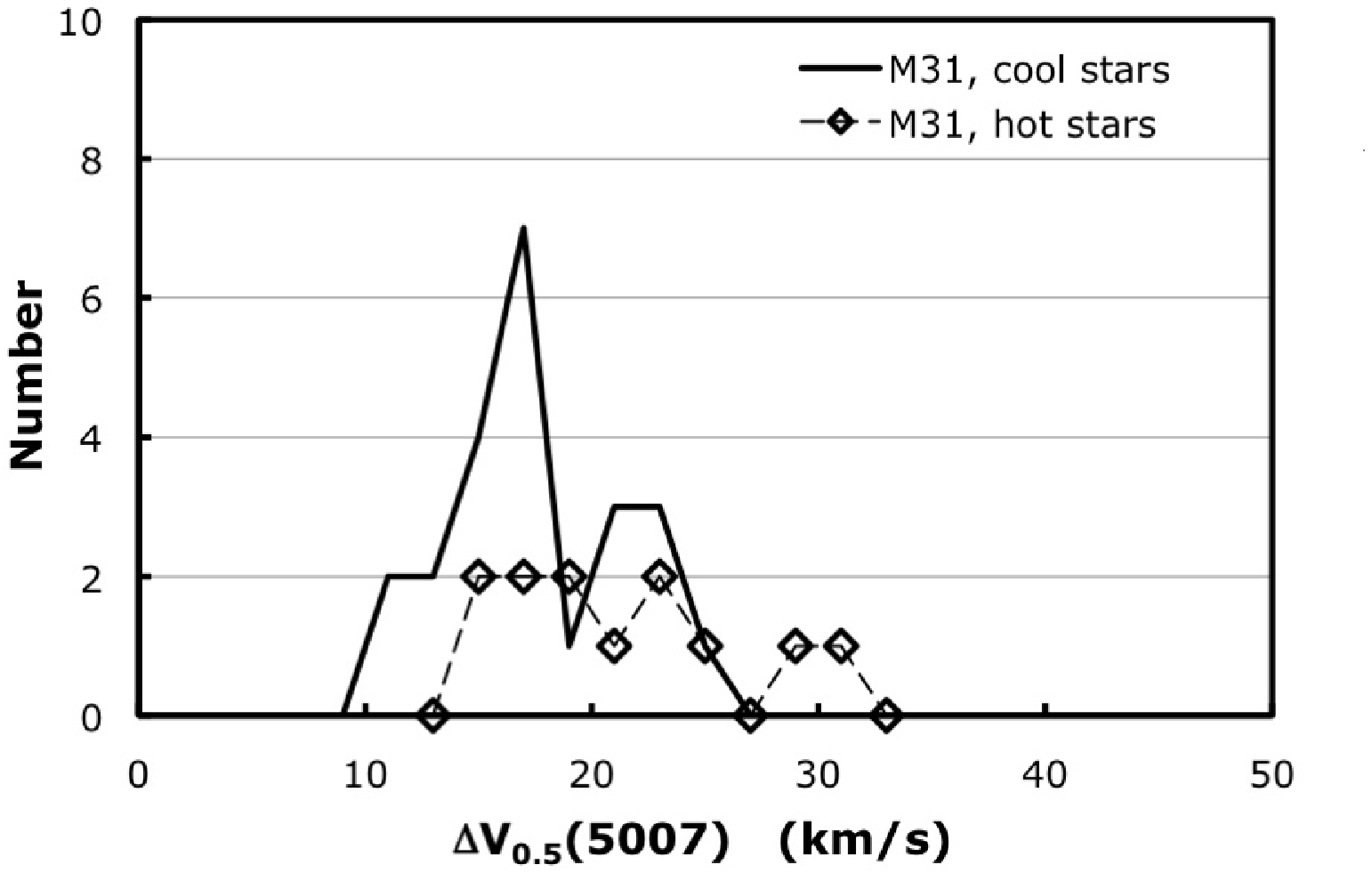} 
 \includegraphics[width=0.49\columnwidth]{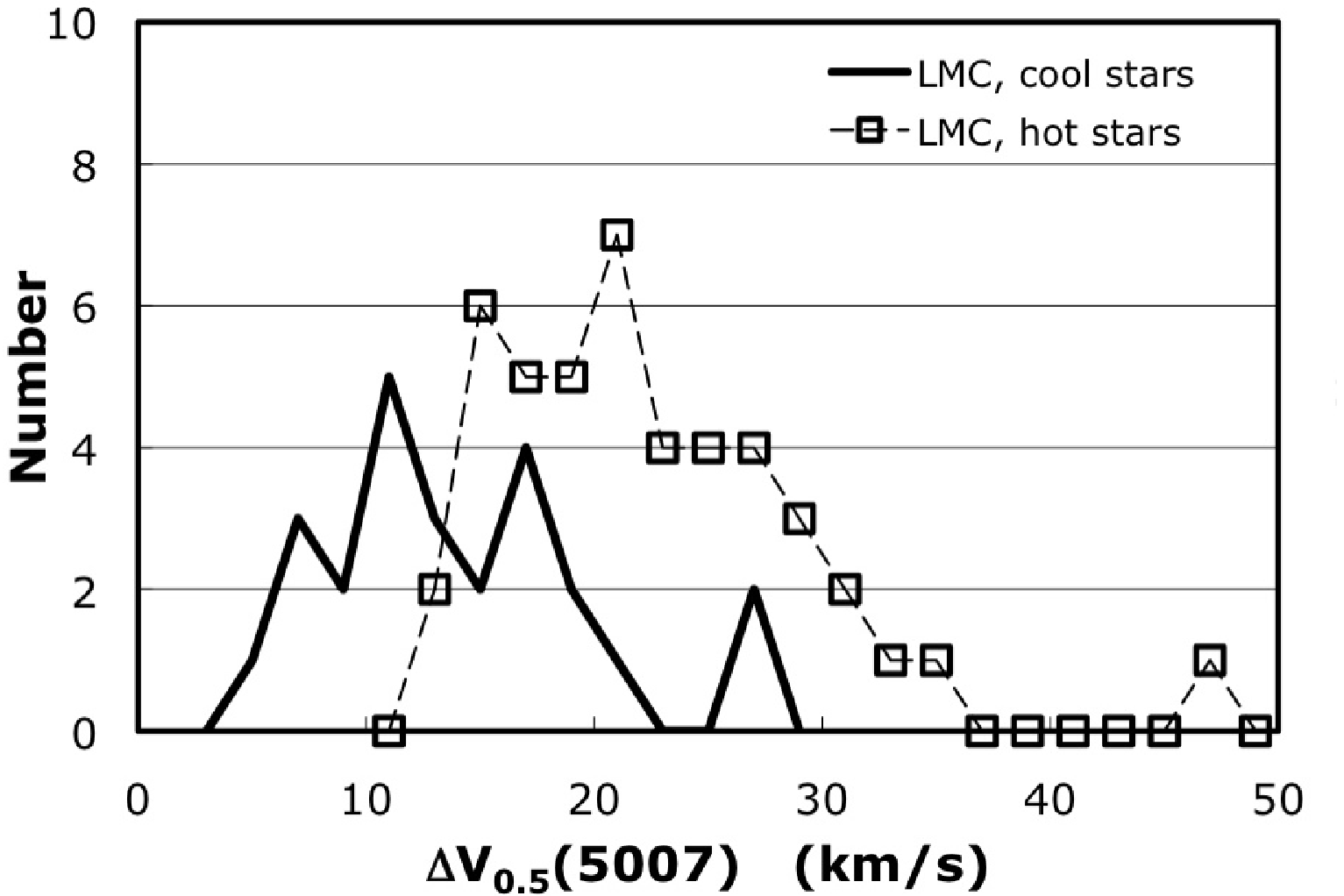} 
 \includegraphics[width=0.49\columnwidth]{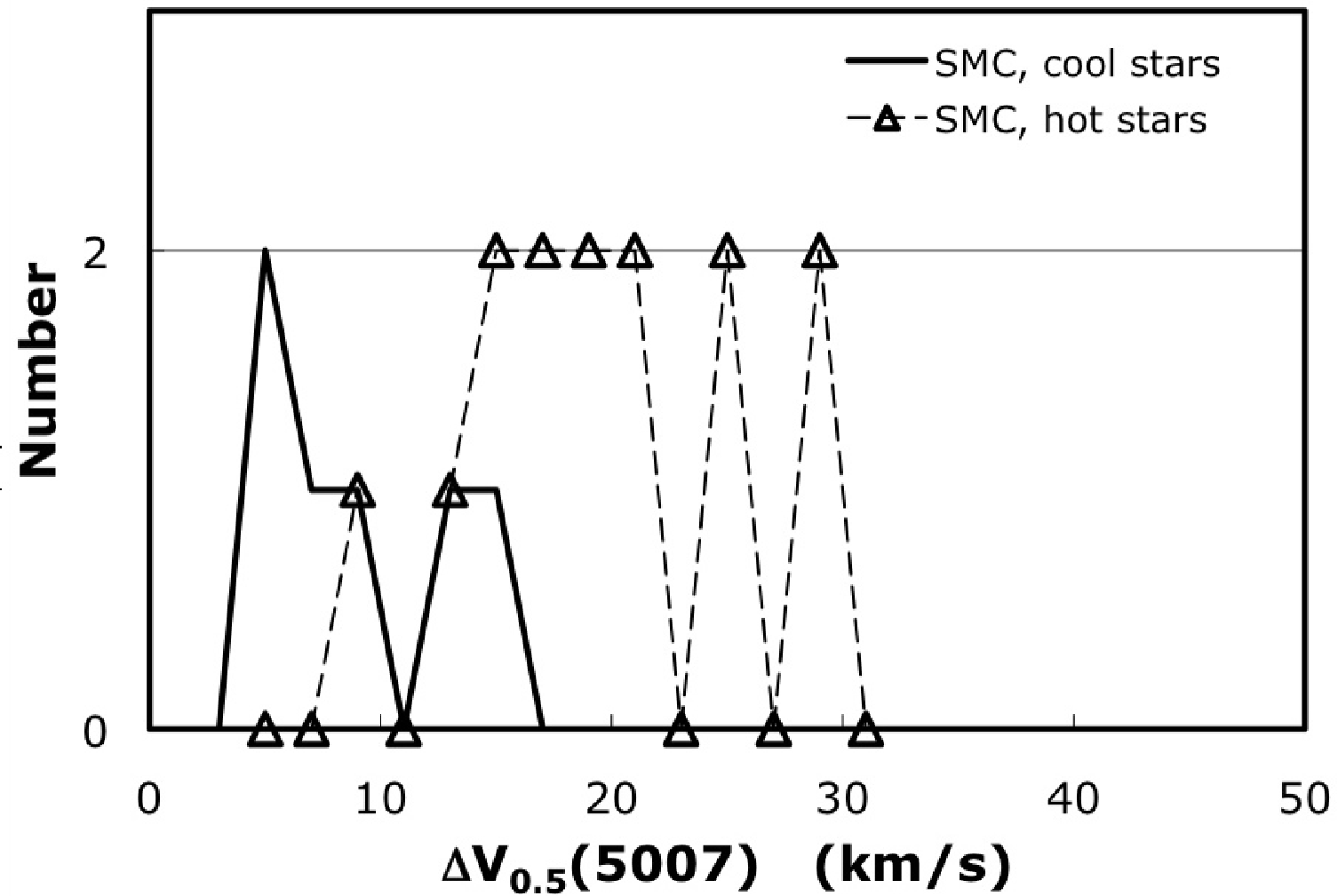} 
 \caption{The line width distributions in the light of [O~{\sc iii}]$\lambda$5007 are shown for the bright planetary nebulae in the Milky Way bulge, the bulge of M31, the LMC, and the SMC.  In all cases, the planetary nebulae are divided into subsamples with cool and hot central stars, with hot central stars being those whose nebulae have an intensity $I($He~{\sc ii} $\lambda 4686)/I(\mathrm H\beta) > 0.1$.  In all cases, the line width distribution for planetary nebulae with hot central stars is shifted to higher velocities.  }
   \label{fig_time_evol}
\end{center}
\end{figure}

From Fig. \ref{fig_time_evol}, it is clear that the line widths measured for bright planetary nebulae in the bulges of M31 and the Milky Way and in the Magellanic Clouds all increase for planetary nebulae surrounding more evolved central stars.  \cite[Richer et al. (2010a)]{richeretal2010a} found a similar result for samples of planetary nebulae from the Milky Way bulge spanning a larger range of evolutionary states.  Therefore, it seems clear that the acceleration of the nebular shell during the central star's evolution to higher temperature is a general phenomenon.  

The evolution of the internal kinematics after the cessation of nuclear burning, while the central star fades to become a white dwarf, is less clear.  The most extensive study to date of evolved planetary nebulae in the Milky Way disc (\cite{pereyraetal2011}) find that the objects containing the most luminous central stars, close to the point at which nuclear reactions are extinguished, have the largest expansion velocities.  The interpretation of this result, however, is not completely clear, as it might arise as a result of (a) backflow if the hot bubble is significantly deflated, (b) seeing matter interior to that seen at earlier phases because of the central star's lower luminosity, or (c) deceleration due to the accumulation of material from the interstellar medium.

\begin{figure}[t]
\begin{center}
 \includegraphics[width=0.495\columnwidth]{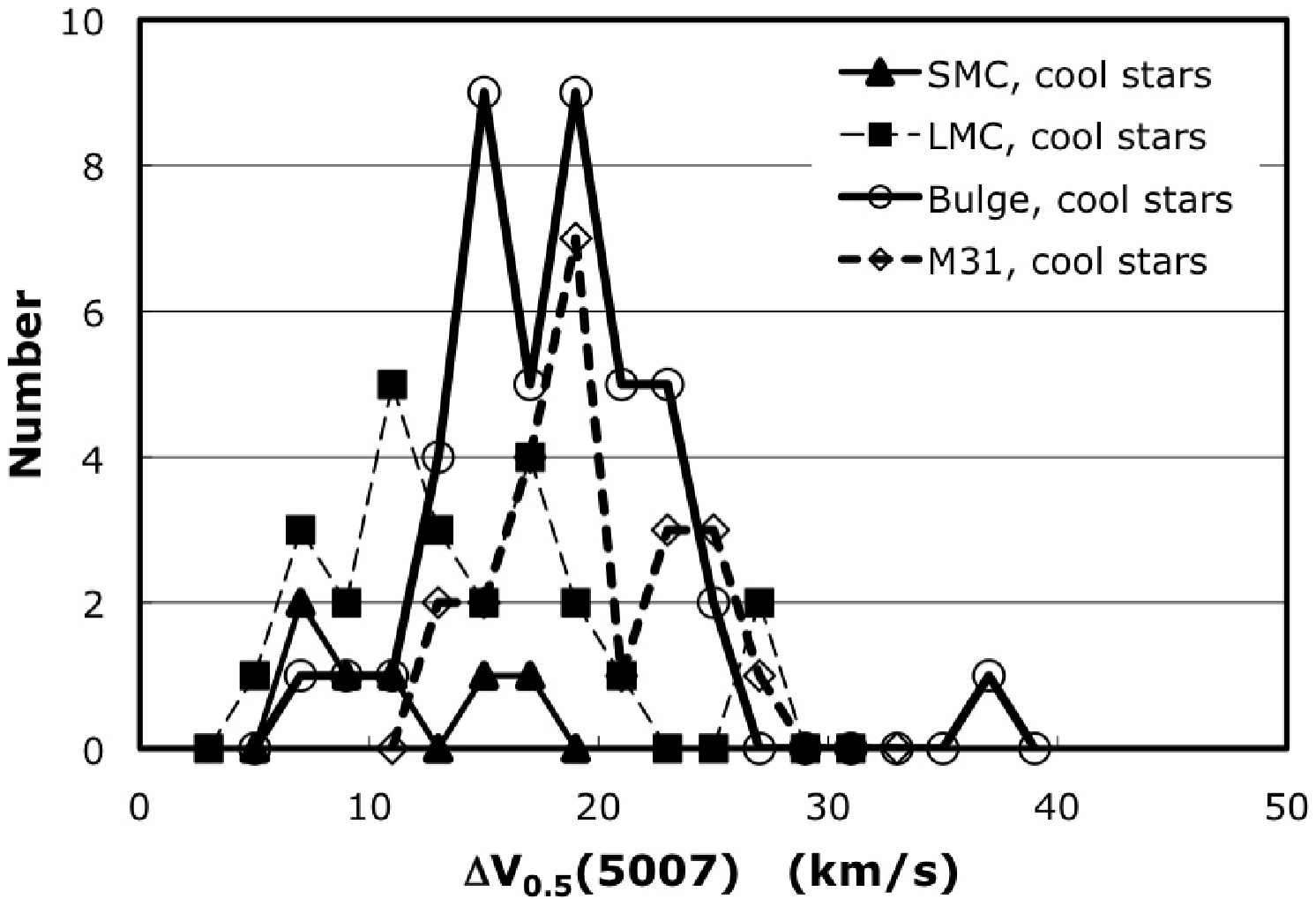} 
 \includegraphics[width=0.485\columnwidth]{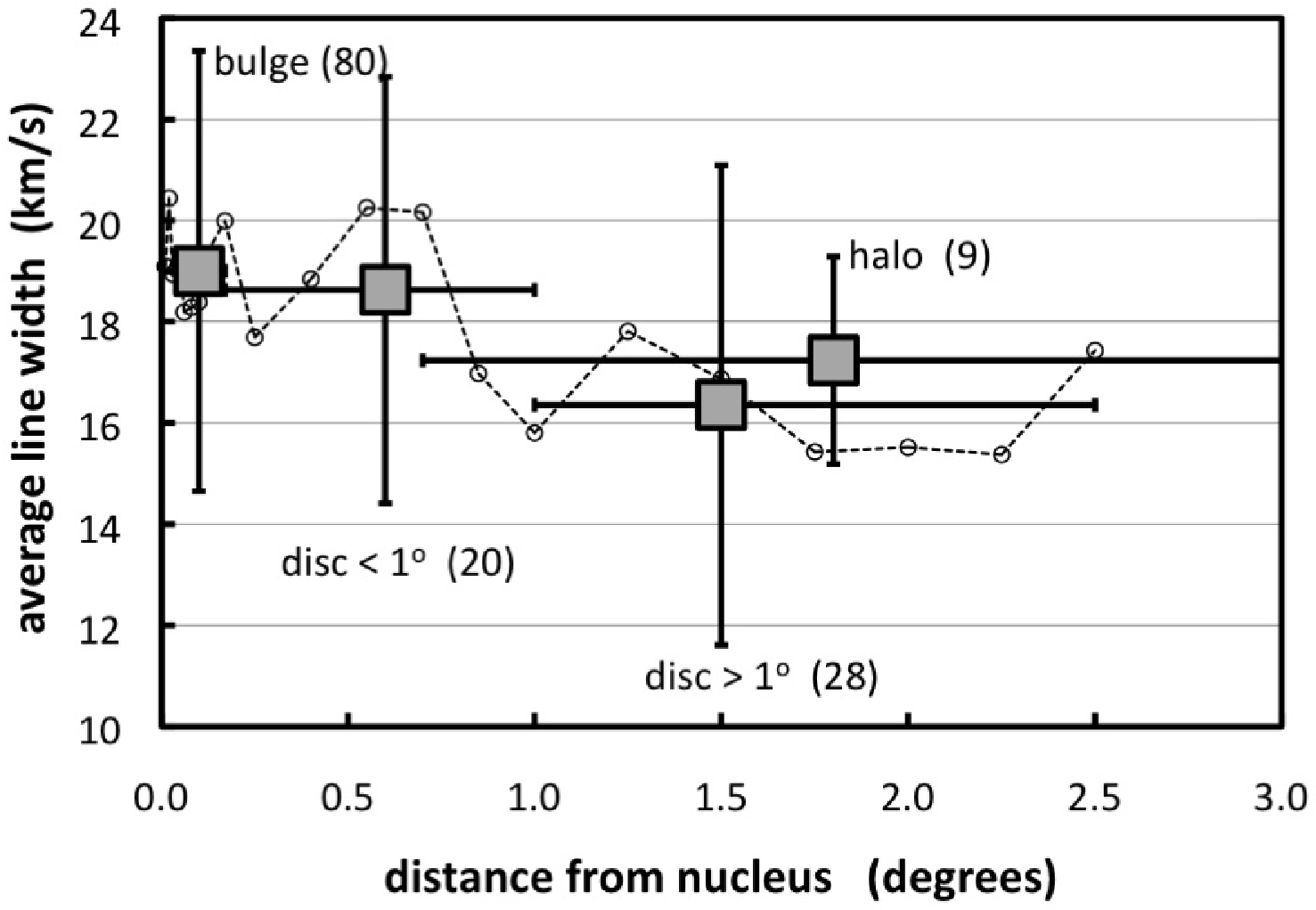} 
 \caption{The effect of metallicity upon the evolution of intrinsically bright planetary nebulae is slight.  The left panel compares the line width distributions for planetary nebulae containing cool central stars (defined as in Fig. \ref{fig_time_evol}) in the Magellanic Clouds and the bulges of M31 and the Milky Way.  The latter two line width distributions are clearly shifted to higher velocities.  The right panel presents the mean velocity observed for planetary nebulae in M31 as a function of the projected distance from the nucleus.  Different samples are shown as well as a running mean (see \cite{richeretal2010b} for details).  The mean clearly decreases as a function of radial distance.  Over the same range, both abundances and the pressure of the interstellar medium decrease as well, suggesting that the trend is due to metallicity (a drop in pressure should allow greater line widths, \emph{if} the environment has a significant effect).}
   \label{fig_metallicity}
\end{center}
\end{figure}

As deduced from Fig. \ref{fig_linewidth_galaxies}, the effect of metallicity is slight among intrinsically bright planetary nebulae.  However, it is probably not irrelevant.  In Fig. \ref{fig_metallicity}, it is clear that bright planetary nebulae with cool central stars in the bulges of M31 and the Milky Way have a line width distributions shifted to higher values than do their counterparts in the Magellanic Clouds.  For bright planetary nebulae with hot central stars, this difference is much less evident (not shown), though there is a slight shift between the distributions for objects in the Milky Way (higher values) compared to those in the LMC, which are the only samples large enough for the difference to be relevant.  The bright planetary nebulae in the Magellanic Clouds have lower metallicities than their counterparts in the bulges of M31 and the Milky Way, and they may also have more massive progenitor stars.  The models of \cite[Sch\"onberner et al. (2010)]{schonberneretal2010} indicate that the nebular shells are accelerated more at lower metallicity, but their models use the same density and velocity structure for the AGB envelope at all metallicities.  While this has the virtue of isolating the effects of greater thermal pressure and lower wind energy at lower metallicity, the result may be misleading if nature really varies the initial condition (AGB envelope structure).  Considering lower AGB wind velocities at lower metallicity could account for the difference in line width distributions for planetary nebulae with cool central stars, and allow the greater thermal pressure at lower metallicity to eventually erase the difference at later times (hot central stars).  

From the right panel of Fig. \ref{fig_metallicity}, it is also clear that the mean line width decreases with radius in M31.  The difference between the bulge and outer disk is statistically significant (\cite{richeretal2010b}).  While it might be tempting to argue that this could be due to the pressure of the interstellar medium in the disk, over the range from $0.5^{\circ}$ to $1.5^{\circ}$, the pressure decreases by an order of magnitude (\cite{braunetal2009}).  So, if the environment had any significant effect upon the expansion of these bright planetary nebulae, those at larger radii should have greater line widths, but the opposite is observed.  On the other hand, the metallicity decreases by a factor of 3 over the radius range from $0.5^{\circ}$ to $2^{\circ}$ (\cite{vennetal2000}), so metallicity could also explain this trend.   

So far, the age of the progenitor stellar population appears to have little effect.  However, our ability to discriminate the mass of stellar progenitors is limited to studying chemical abundances (e.g., \cite{richermccall2008}), which isn't a very sensitive technique.  At present, such studies imply little variation in progenitor masses in general, though rare exceptions do exist (and are not necessarily well-understood).  

\section{Conclusions}

The kinematics of intrinsically bright planetary nebulae do not depend sensitively upon environment, metallicity, or the presence of ongoing star formation.  The internal kinematics of bright planetary do, however, depend upon the evolutionary stage (time).  There also appears to be a slight dependence upon metallicity.  The time evolution of the kinematics established for the Magellanic Clouds (\cite{dopitaetal1985, dopitaetal1988}) are extended to the bulges of M31 and the Milky Way, and presumably all environments, as hydrodynamical models predict (e.g., \cite{kahnbreitschwerdt1990, villaveretal2002, schonberneretal2010}).  Therefore, a \lq\lq typical" expansion velocity should not be quoted for a planetary nebula generally (often done), but instead for a give stage of evolution.  Metallicity probably introduces several competing effects that largely cancel each other out: lower AGB wind velocity and weaker wind from the central star versus greater thermal pressure in the nebular shell.  

It is a pleasure to acknowledge the helpful discussions with many colleagues, particularly those in Ensenada.  The support of grants from CONACyT (82066) and DGAPA-UNAM (110011) are greatly appreciated.

\end{document}